\documentclass{SCIS2018}
\usepackage{graphicx}

\begin{document}

\ArticleType{INVITED REVIEW}
\Year{2018}
\Month{}
\Vol{61}
\No{}
\DOI{}
\ArtNo{}
\ReceiveDate{}
\ReviseDate{}
\AcceptDate{}
\OnlineDate{}

\title{AI for 5G: Research Directions and Paradigms}{AI for 5G: Research Directions and Paradigms}


\author[1]{Xiaohu YOU}{}
\author[1]{Chuan ZHANG}{{chzhang@seu.edu.cn}}
\author[1]{Xiaosi TAN}{}
\author[1]{Shi JIN}{}
\author[2]{Hequan WU}{}

\AuthorMark{YOU X H}

\AuthorCitation{YOU X H, ZHANG C, TAN X S, et al}


\address[1]{National Mobile Communications Research Laboratory, Southeast University, Nanjing {\rm 210096}, China}
\address[2]{Chinese Academy of Engineering, Beijing {\rm 100088}, China}

\abstract{The 5th wireless communication (5G) techniques not only fulfil the requirement of $1,000$ times increase of internet traffic in the next decade, but also offer the underlying technologies to the entire industry and ecology for internet of everything. Compared to the existing mobile communication techniques, 5G techniques are more-widely applicable and the corresponding system design is more complicated. The resurgence of artificial intelligence (AI) techniques offers as an alternative option, which is possibly superior over traditional ideas and performance. Typical and potential research directions to which AI can make promising contributions need to be identified, evaluated, and investigated. To this end, this overview paper first combs through several promising research directions of AI for 5G, based on the understanding of the 5G key techniques. Also, the paper devotes itself in providing design paradigms including 5G network optimization, optimal resource allocation, 5G physical layer unified acceleration, end-to-end physical layer joint optimization, and so on.}

\keywords{5G mobile communication, AI techniques, network optimization, resource allocation, unified acceleration, end-to-end joint optimization}

\maketitle

\section{Introduction}
\label{sec:intro}

The fifth generation mobile network (5G) applies the next generation of mobile telecommunication standards which targets at the needs of 2020 and beyond.
It aims at providing a complete wireless communication system with diverse applications.
Specifically, 5G is going to be responsible for supporting three generic services, which are classified as enhanced mobile broadband (eMBB), massive machine-type communications (mMTC) and ultra-reliable and low-latency communications (URLLC) (also referred to as mission-critical communications). These applications suggest new performance criteria for latency, reliability, connection and capacity density, system spectral efficiency, energy efficiency and peak throughput that need to be addressed with the 5G technology. To meet these criteria, ongoing researches are conducted in many areas, mainly focused on key technologies including massive multiple-input multiple-output (MIMO), new radio access technology (RAT), heterogeneous ultra-densification networks (UDN), channel coding and decoding (e.g. polar codes) and mmWave access~\cite{1}. Also, 5G networks will inevitably be heterogeneous, with multiple modes and devices involved through one unified air interface tailored for specific applications. Hence, architectures as the dense HetNet are involved and 5G systems are going to be virtualised and implemented over cloud data centers. Network slicing will be a major feature of a 5G network, as will the use of a new air interface designed to dynamically optimize the allocation of network resources and utilize the spectrum efficiently.

The 5G technology standards are in development, getting complete and mature~\cite{2,3}.
In December 2017, 3G Partnership Project (3GPP) officially announced the new standards for 5G New Radio (NR) which include supports for 5G Non-Standalone architecture (NSA) and eMBB~\cite{4}. On June 14, 2018, 3GPP formally completed the Standalone (SA) version of the 5G NR standard, marking a long-awaited target date for 5G standardization~\cite{5}. These announced standards effectively set the stage to launch full-scale and cost-effective development of 5G networks.
Compared to the current 4G networks, 5G NR:
(1) enhances the MIMO systems with the massive MIMO technology;
(2) makes completion to the time slot structure and resource block (RB) allocation of the orthogonal frequency-division multiplexing (OFDM), proposing a more flexible air interface;
(3) will introduce the non-orthogonal multiple access (NOMA) to support the Internet of Things (IoT) in the near future;
(4) follows the previous distributed antenna systems~\cite{6}, splits the wireless functions into distributed units (DU) and central units (CU) and applies network virtualization and network slicing techniques based on cloud computing.

Overall, 5G networks will tailor the provisioning mechanisms for more applications and services, hence is more challenging with the complicated configuration issues and evolving service requirements.
Before 5G, researches of communication systems mainly aim at satisfactory data transmission rate and supportive mobility management. In the 5G era, the communication systems will gain the abilities to interact with the environment, and the targets are expanded to joint optimizations of ever-increasing numbers of key performance indicators (KPIs) including latency, reliability, connection density, user experience, etc~\cite{7}.
Meanwhile, new features like the dynamic air interface, virtualised network and network slicing introduce complicated system design and optimization requirements as long as challenges for network operation and maintenance.
Fortunately, such problems falls into the field
of artificial intelligence (AI) which provides brand new concepts and possibilities beyond traditional methods, hence has recently regained attention in the field of communications in both academia and industry.

AI is dedicated to allowing machines and systems to function with intelligence similar to humans.
The field of AI research was born in 1950s, which experienced ups and downs, and is revived in recent years due to the rapid development of modern computing and data storage technologies.
The general problem of simulating intelligence contains sub-problems like reasoning, inference, data fitting, clustering and optimization, which involves approaches including genetic algorithms~\cite{8} and artificial neural networks (ANN)~\cite{9,10}.
Specifically, AI learning techniques have constructed a universal framework for various problems and have made tremendous progress, becoming the state of the art in different fields.

AI learning tasks are typically classified into two broad categories, supervised and unsupervised learning, depending on whether labels of the training data are available to the learning system. And another learning approach, reinforcement learning, is not exactly supervised neither unsupervised, hence can be listed in a new category.

\begin{figure}[htbp]
\centering
\includegraphics[width=4.5in]{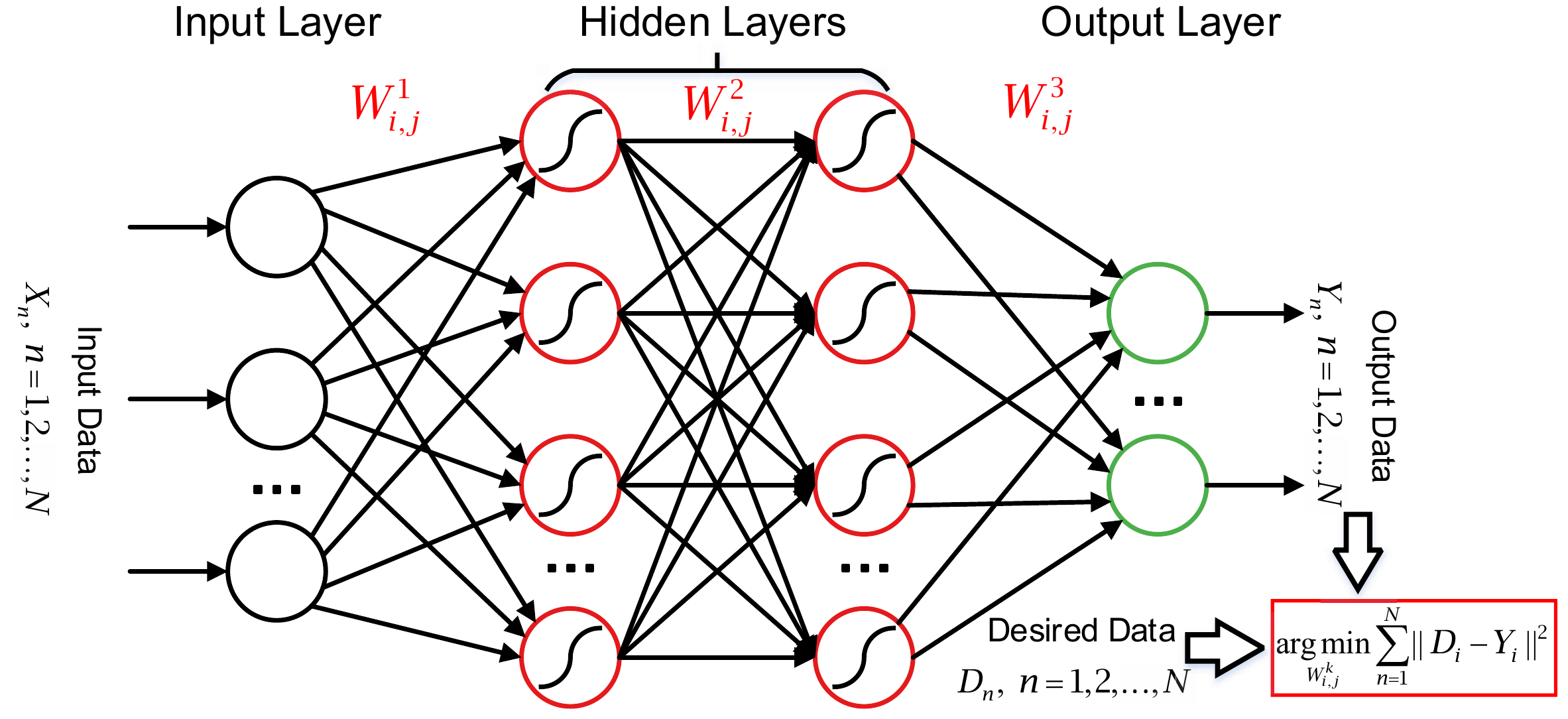}
\caption{Example of supervised learning: learning in deep neural networks.}
\label{fig:DNN}
\end{figure}

\begin{figure}[htbp]
\centering
\includegraphics[width=4in]{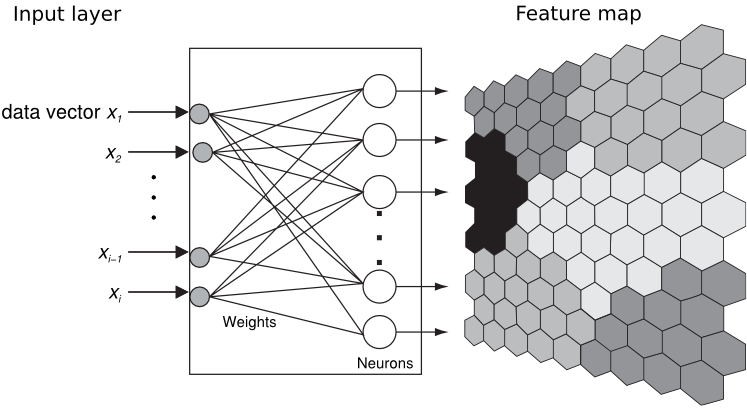}
\caption{Example of unsupervised learning: self organizing map.}
\label{fig:som}
\end{figure}

\begin{figure}[!t]
\centering
\includegraphics[width=4.5in]{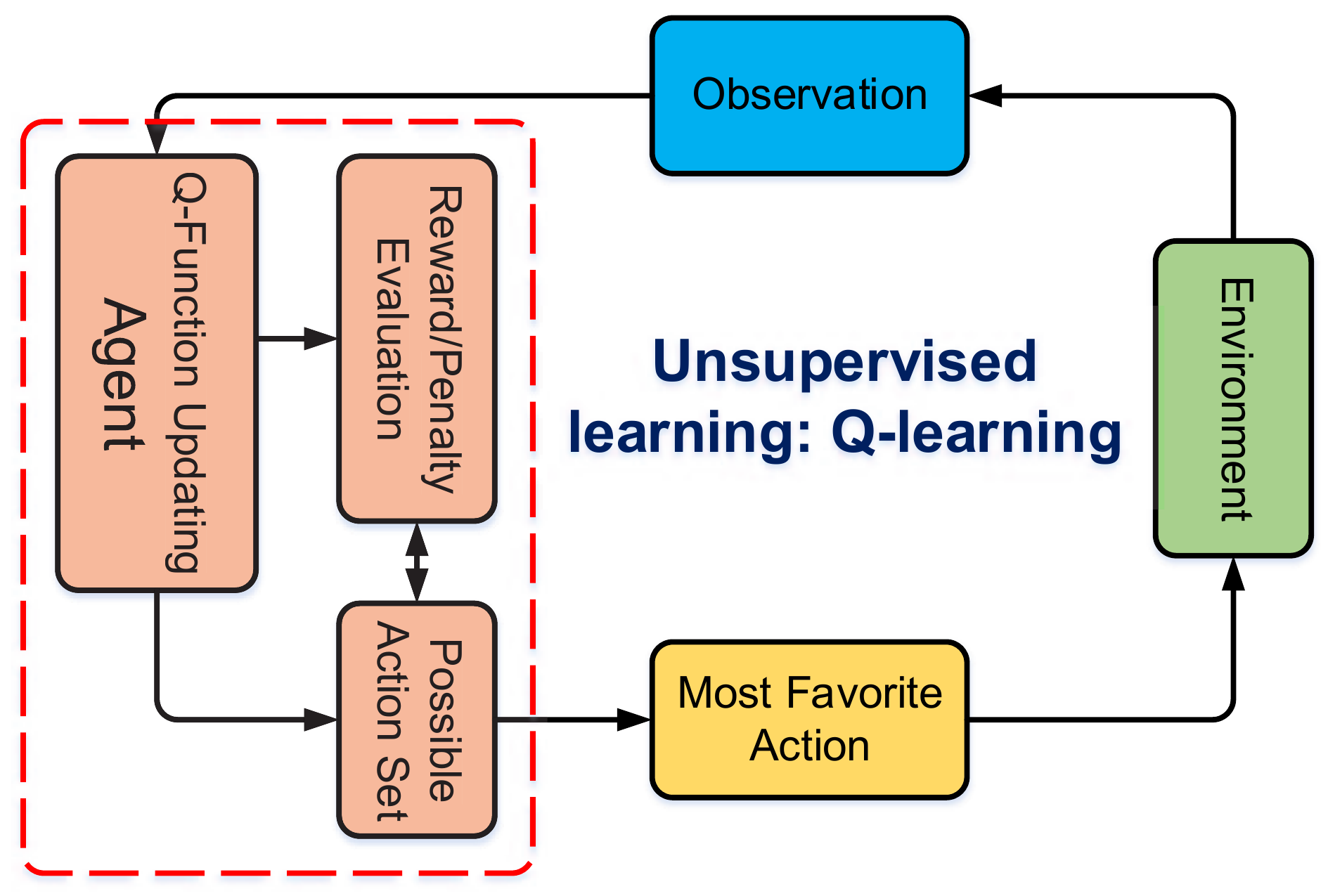}
\caption{Example of reinforcement learning: Q-learning.}
\label{fig:qlearning}
\end{figure}

\begin{itemize}
\item \textbf{Supervised learning:} Sample data pairs of inputs and desired outputs are feed into the computer, and the goal is to learn a general function that relates the inputs to the outputs and further detects the unknown outputs of the future inputs. One typical example of supervised learning is illustrated in Figure \ref{fig:DNN}, in which labeled data pairs are feed in a multi-layer deep neural network (DNN) to train the weights between the nodes in the DNN. The training is carried out offline, and after convergence, the trained DNN will be ready for recognition and inference of new inputs.
\item \textbf{Unsupervised learning:} In unsupervised learning, no labels are given to the learning algorithm and structure in its input should be found on its own. Self-organizing map (SOM) is an example that is trained using unsupervised learning. In SOM, unlabeled data are feed in to a neural network to produce a low-dimensional (usually two-dimensional), discretized representation of the input space of the training samples, called a map (as illustrated in Figure \ref{fig:som}) , and is therefore a method to do dimensionality reduction.
\item \textbf{Reinforcement learning:} This technique is based on alternative interaction between ``Agent" and ``Environment" and the process is illustrated in Figure \ref{fig:qlearning}. The ``Agent" will perform certain action and as a result of this action his state will change which leads to either a reward or a penalty. The Agent will then decide the next action based on this result. By iterating through action and reward/penalty process, Agent learns the Environment.

\end{itemize}

Popular learning methods in AI learning problems includes:
\begin{itemize}
\item \textbf{Backpropagation (BP):} Backpropagation is a method used in ANN which is in the category of gradient descent~\cite{9}. It iteratively calculates gradient of the defined loss function with respect to the weights in the ANN to finally make the output of the ANN close to the known training label. The dynamic learning rate optimization and acceleration of the BP learning are introduced in the author's previous works~\cite{9.5,10}. In ~\cite{11}, local minima of the BP surface is discussed. Recently, BP is commonly used to train deep neural networks (DNN), a term referring to neural networks with more than one hidden layer. For example, convolutional neural network (CNN) is a class of feed-forward DNN, which has multiple hidden layers including convolutional layer, pooling layer, fully-connected layer and ReLU layer. CNN can be trained with the BP method efficiently, especially in the fields of image and voice recognition.
\item \textbf{Q-learning:} The Q-learning algorithm is also referred to as Bellman's algorithm~\cite{12}, which is a classical algorithm for the reinforcement learning. In this algorithm, a function (Q function) is defined to evaluate the actions of the ``Agent" based on the current ``Environment" and output the result in the form of award or penalty. At a certain step, all the possible actions of the ``Agent" will be evaluated by the Q function, and the action with the maximum award in the current ``Environment" will be selected as the next step and will be actually carried out.
    Bellman proposed a Bellman equation which is a recursive expression that relates the Q functions of consecutive time steps. Bellman equation basically allows us to iteratively update and approximate the Q function through online temporal difference learning. More details can be found in~\cite{13}.
\end{itemize}

\section{Research directions for AI in 5G}
\label{sec:ai45g}

As a universal intelligent problem-solving technique, AI can be broadly applied in the design, configuration and optimization for the 5G networks. Specifically, AI is relevant to three main technical problems in 5G:
\begin{itemize}
\item \textbf{Combinatorial optimization:} One typical example of the combinatorial optimization problem in 5G NR is the network resource allocation. Given a resource-limited network, an optimized scheme should be figured out to allocate resources to different users who share the network such that the utilization of the resource achieves maximum efficiency. As the application of the HetNet architecture in 5G NR with features like network virtualization, network slicing and self-organizing networks (SON), the related network resource allocation problems are becoming more complicated, which requires more effective solutions.
\item \textbf{Detection:} The design of the communication receiver is an example of the detection problem. An optimized receiver is able to recover the transmitted messages based on the received signals, achieving minimized detection error rate. Detection will be challenging in 5G within the massive MIMO framework.
\item \textbf{Estimation:} The typical example is the channel estimation problem. 5G requires accurate estimation of the channel state information (CSI) to achieve communications in spatially correlated channels of massive MIMO. The popular approach is the so-called training sequence (or pilot sequence), where a known signal is transmitted and the CSI is estimated using the combined knowledge of the transmitted and received signal.
\end{itemize}

Many researches have already been done for the application of AI in 5G as in literatures~\cite{14,15,16,17,18,19,20,21,22}. However, due to the limitations of both the communication systems and AI, some of the applications may be restricted. Firstly, after years of research and test, conventional methods have shown their abilities to handle the communication systems. A complete framework with conventional techniques have been formed, which is effective, mature and easy to implement for real world scenarios. Secondly, the capacity of a communication system is constrained with certain upper bounds (e.g., the Shannon limit), and some of the well-designed methods can reach near-optimal performance suffering negligible loss with respect to the capacity bound. For example, in~\cite{23} a transmitter optimization method for MIMO is proposed based on an iterative water-filling algorithm, which closely achieves near-Shannon performance in the general jointly correlated MIMO channels. This kind of methods will not be over-performed by even the most-advanced AI technique. Moreover, there're still obstacles to apply AI learning in real-world problems due to the convergence issues for training. Careful checks should be done to make sure the optimal performance can be ``learned" with AI in every specific problem in the communication system. Finally, AI algorithms usually bring large computational complexity, which makes them not that competitive compared to conventional methods if the performance improvement is minor.

With all these limitations, nonetheless, AI still has great potentials and prospects in the communication system of the 5G era. As introduced above, 5G brings complicated configuration issues and evolving service requirements, resulting in new problems that are hard to be modeled, solved or implemented within the current conventional framework. Hence, new opportunities as long as challenges are brought by 5G for AI techniques. From all the challenging problems in 5G, typical and potential research directions to which AI can make promising contributions need to be identified, evaluated, and investigated.

In this paper, we summarize the potential application directions for AI in 5G in four main categories:
(1) Problems difficult to model;
(2) Problems difficult to solve;
(3) Uniform implementation;
(4) Joint optimization and detection.
In the following analysis and examples, we will see that for problems in (1) and (2), conventional methods are barely effective and AI techniques are expected to be promising; on the other hand, for (3) and (4), the potential of AI is problem-dependent compared to conventional methods, and careful investigation should be done to identify if AI is beneficial.
\begin{itemize}
\item \textbf{Problems difficult to model:} The network optimization problems in communication systems is overall a kind of technical problem that is hard to model, which includes typical issues such as the network coverage, interference, neighboring cell selection and handover. Current solutions mostly depend on experience of the engineers. For 5G NR scenarios, these problems are more challenging due to the complicated network structures and large number of KPIs. Applications of new features like massive MIMO beamforming~\cite{24} are associated with high-dimensional optimization parameters and the optimization problem itself is difficult to model. Also, 5G NR involves multiple KPIs including peak data rate, spectral efficiency, latency, connection density, quality of experience (QoE) and so on. These KPIs should be jointly optimized even if some of them contradicts with each other~\cite{7}. In these situations, an overall optimization model cannot be achieved with conventional methods and AI techniques are expected to be able to handle them.

\item \textbf{Problems difficult to solve:} Network resource allocation is a key issue in 5G NR~\cite{24,25},
    which includes specific issues in inter-cell resource block allocation, orthogonal pilot resource allocation,
    beamforming resource allocation, massive MIMO user clustering and resource pool deployment in virtualized networks.
    The network resource allocation aims at maximizing the throughput of the network while balancing the service rate. It's mostly a NP-hard combinatorial optimization problem, and the computational complexity to solve this kind of problem increases exponentially as the size of the systems.
    Traditional solutions use static partition of the network to cut down the computational cost for a sub-optimal solution. Nowadays, with the assistance of modern computing technologies, AI will be an new effective solution for these problems.

\item \textbf{Uniform implementation:} Conventional methods are designed in a divide-and-conquer manner for some function blocks in 5G NR. For example, the physical layer in 5G NR consists of a series of signal processing blocks such as multiuser MIMO space-time processing, NOMA signal detection and encoding and decoding for LDPC or polar codes. Researchers have attempted to optimize the algorithms and implementations of each processing module and achieved success in practice. However, the efficient and scalable implementation of the entire communication system is lacking, with guaranteed performance. It is noted that, AI techniques are supposed to be capable for handling each of the modules~\cite{14,15,16,17,18,19,20,21,22}. This inspires us to further develop a uniform AI-based implementation which works jointly for all the key modules in the 5G NR physical layer~\cite{26}. By unifying the modules with AI methods in both algorithm and hardware, the design, configuration and implementation of the physical layer communications will be simpler, faster, more economical and more efficient.

\item \textbf{Joint optimization and detection:}
    An intuitive idea for applying AI in 5G is to simply substitute the conventional modules of transmitters and receivers by ANN. However, the capacity of the channels is bounded by the Shannon limit and the improvement by using ANN is constricted. Also as discussed above, the complexity and the convergence of training should be carefully examined in this area. Compared to this intuitive method, AI is more potential in a bigger picture of the cross-layer joint optimization problem which cannot be solved efficiently with conventional methods~\cite{27}.
    Typical examples include the joint optimizations for the physical and media access control layers~\cite{28}, joint source and channel optimizations~\cite{29} and the joint optimization for algorithm and hardware implementation~\cite{30}.
\end{itemize}

\section{Paradigms of AI in 5G}
\label{sec:example}

In this section, examples of the application of AI techniques in 5G are presented,
which cover four different problems in 5G including: network resource allocation,
SON, uniform 5G accelerator and the optimization of the end-to-end physical layer communication.

\subsection{AI for SON: Automatic root cause analysis}
Self-organizing networks (SONs) establish a new concept of network management which provide intelligence
in the operation and maintenance of the network. It has been introduced by 3GPP as a key component of LTE network.
In the 5G era, network densification and dynamic resource allocation will result in new problems for
coordination, configuration and management of the network, hence bringing increased demand for the improvements
of the SON functions. SON modules in mobile networks can be divided into three main categories:
self-configuration, self-optimization and self-healing.
The main objectives of SON are to automatically perform network planning, configuration and optimization without
human intervention, in order to reduce the overall complexity, Operational Expenditure (OPEX), Capital Expenditure (CAPEX)
and man-made faults. Various researches of AI in SON have been summarized in~\cite{31,32,33}
which includes AI applied in automatic base station configuration, new cell and spectrum deployment,
coverage and capacity optimization, cell outage detection and compensation, etc.,
using approaches including ANN, ant colony optimization, genetic algorithm, etc..

In this section, we introduce the automatic root cause analysis framework proposed in~\cite{34} as an example
 for AI in SON. The design of the fault identification system in LTE networks faces two main challenges:
 (1) A huge number of alarms, KPIs and configuration parameters can be taken as fault indicators in the system.
 Meanwhile, most of the symptoms of these indicators are not labeled with fault causes, hence are difficult to identify;
 (2) The system is not automatic and experts are involved to analyze each fault cause.
 With the huge amount of high-dimensional data, human intervention is not efficient while expensive.
 Authors of~\cite{34} proposes an AI-based automatic root cause analysis system which combines
 supervised and unsupervised learning techniques as summarized in the following steps:

\begin{itemize}

\item[1.] Unsupervised SOM training.
SOM as shown in Figure \ref{fig:som} is applied for an initial classification of the high-dimensional KPIs.
An SOM is a type of unsupervised neural network capable of acquiring knowledge and learning from a set of unlabeled data.
It will process high-dimensional data and reducing it to a two-dimensional map of neurons that preserves
the topological properties of the input data.
Hence, inputs close to each other will be mapped to adjacent neurons.
By this unsupervised process, the high-dimensional KPIs are mapped into a lower-dimensional map
which can classify new KPI data by finding the closest neurons.

\item[2.] Unsupervised clustering.
After SOM training, all the neurons in the SOM system will be clustered into a certain number of groups
using an unsupervised algorithm. Since the SOM neurons are already ordered and the difference between
the original inputs can be represented by Euclidean distance between the corresponding neurons,
clustering algorithms based on Euclidean distance, e.g. Ward’s hierarchical method,
can be applied to cluster the neurons.

\item[3.] Labeling by Experts.
After the above two steps, the original high-dimensional data are clustered into several classes.
We will finally have the experts to analyze and identify the fault causes of each obtained cluster
to have all the clusters labeled.
\end{itemize}

\begin{figure}
\centering
\includegraphics[width=4.5in]{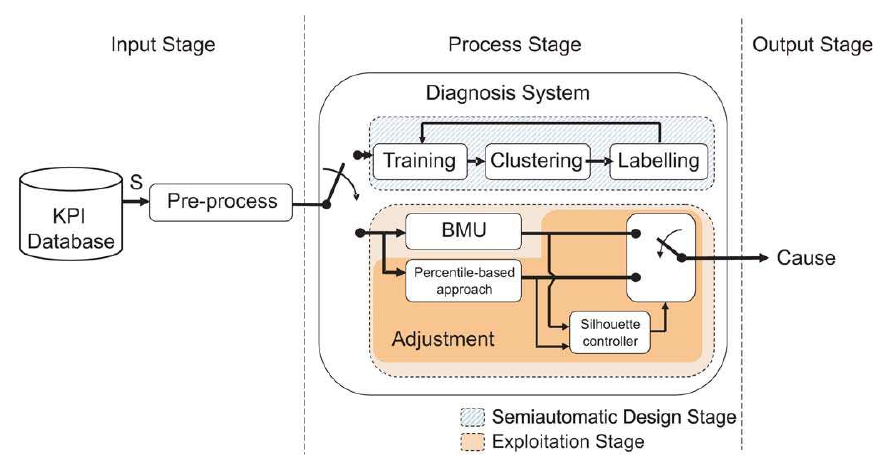}
\caption{Automatic root cause analysis workflow~\cite{34}.}
\label{fig:selfheal}
\end{figure}

With the training, clustering and labeling, an automatic system for network diagnosis is constructed by the workflow
shown in Figure \ref{fig:selfheal}.
For a new input of KPIs, it will firstly be mapped to a neuron in SOM.
Then by the label of the cluster this neuron belongs to, we can identify the fault and the causes.
After obtaining a certain amount of new fault data, we can verify whether the system is right or not
and update it by re-training with the above three steps.
Simulation results presented in~\cite{34} show that the proposed root cause analysis system
is highly accurate even though it's mainly built using unsupervised techniques.

\subsection{AI for resource allocation:  OFDMA downlink resource allocation}

\begin{figure}
\centering
\includegraphics[width=4in]{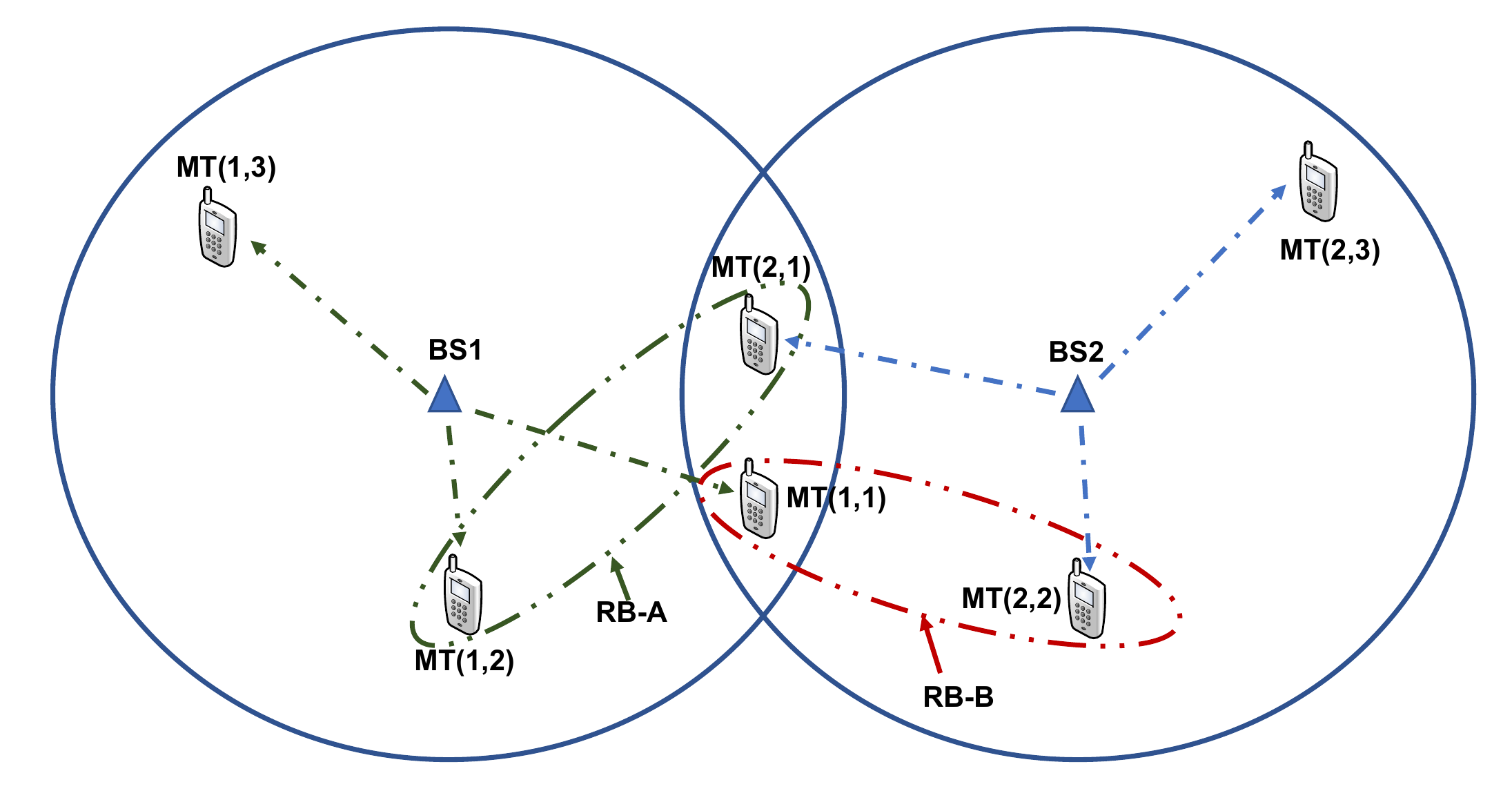}
\caption{Dynamic resource allocation for multi-cell and multi-user systems.}
\label{fig:resourceallocation}
\end{figure}

The OFDM resource block (RB) allocation in 5G NR is more complicated and challenging than ever before
due to the support for the three aforementioned generic services.
In Figure \ref{fig:resourceallocation}, a typical multi-cell, multi-user downlink resource allocation scenario
is illustrated. In this system, the intra-cell interference is eliminated since the RB allocated to
different users in the same cell is orthogonal to each other.
The system interference mainly depends on the inter-cell interference,
which makes the RB allocation for users in neighboring cells important.
Suppose the throughput of each user can be evaluated based on signal-to-interference ratio (SIR),
the target for the optimization of the RB allocation is the maximization of the total system throughput.
This is indeed an NP-hard combinatorial optimization problem with nonlinear constraints.
The complexity of the traditional solution is proportional to the factorial of the number of users in coverage,
which is computationally prohibitive.

Q-learning can be applied to this problem.
Suppose an ``Agent" is in charge of the RB allocation, then the possible ``Action" of this ``Agent",
which is to update the RB for each user, can be selected by the following strategies
as illustrated in Figure \ref{fig:resourceallocation}:
(1) Within the same cell, allocate free RB with higher SIR to the users;
(2) Update the RB allocated to the user with the worst SIR in current cell continuously
to achieve better overall system capacity;
(3) For a certain RB, pair or cluster the user with worst SIR in the current cell
with users with the best SIR from the neighboring cell.
The first two strategies are intuitive.
The third one is applied to avoid allocating the same RB to users in neighboring cells
that are located close to the boarder, since in such situations the involved users
cannot acquire essential SIR to work properly regardless of the transmitting power of the base stations.

After defining all the possible ``Actions", the ``Agent" will evaluate each of them
to select the next ``Action" to adjust the RB allocation which maximizes the overall capacity of the whole system.
The Q function is updated according to the Bellman equation~\cite{12} at the same time.
We iterate through this process until the Q function converges.

These iterations should also be considered with the optimization of the user power allocation.
In~\cite{35}, a framework based on generalized Nash equilibrium problem (GNEP) is proposed for
the optimization of the power control for users from multiple cells that are assigned the same RB.
Global quality of service (QoS) constraints are also considered hence Lagrange multipliers are introduced
to evaluate the ``Actions" and establish the Q function.

\subsection{AI for baseband signal processing: Uniform 5G accelerator}
 The baseband signal processing in 5G consists of a series of signal processing blocks
 including massive MIMO detection, NOMA detection and decoding for polar codes.
 The increased number of baseband blocks leads to more hardware area and varied implementation structures.
 However, we notice that the belief propagation algorithm based on factor graphs can be applied
 to all the blocks as proposed in~\cite{36,37,38,39,39.5}.
 For each specific block, the frameworks are kept unchanged and we only have to
 adapt the symbol set and constraints of the variables to the certain function.
 Hence, a uniform accelerator for the baseband can be designed based on the belief propagation algorithms
 with configurable variables.

 However, the performance of belief propagation is limited in some baseband blocks in certain scenarios.
 Here, AI can be a possible solution to these problems.
 By improving the belief propagation methods with the AI techniques,
 an AI-based uniform accelerator can be constructed.
 The AI-aided belief propagation algorithms can be designed with the following two methods:
 \begin{itemize}
 \item \textbf{DNN-aided belief propagation:}
 (1) Unfold the factor graph of belief propagation by duplicating the iterations to form a DNN;
 (2) Train the DNN by supervised training.
 Applications of this method in the baseband include the DNN-based polar codes decoder proposed in~\cite{19}
 and the DNN-aided MMO detector proposed in~\cite{40}.

 \item \textbf{Belief propagation-based CNN:}
 (1)Map each node in the factor graph of belief propagation to one pixel in a picture,
 in which connected nodes should be mapped into neighboring pixels;
 (2) Train the CNN using the obtained pictures.
 This method is utilized in the BP-CNN channel decoder proposed in~\cite{41}.
 \end{itemize}

\begin{figure}[!t]
\centering
\includegraphics[width=3.5in]{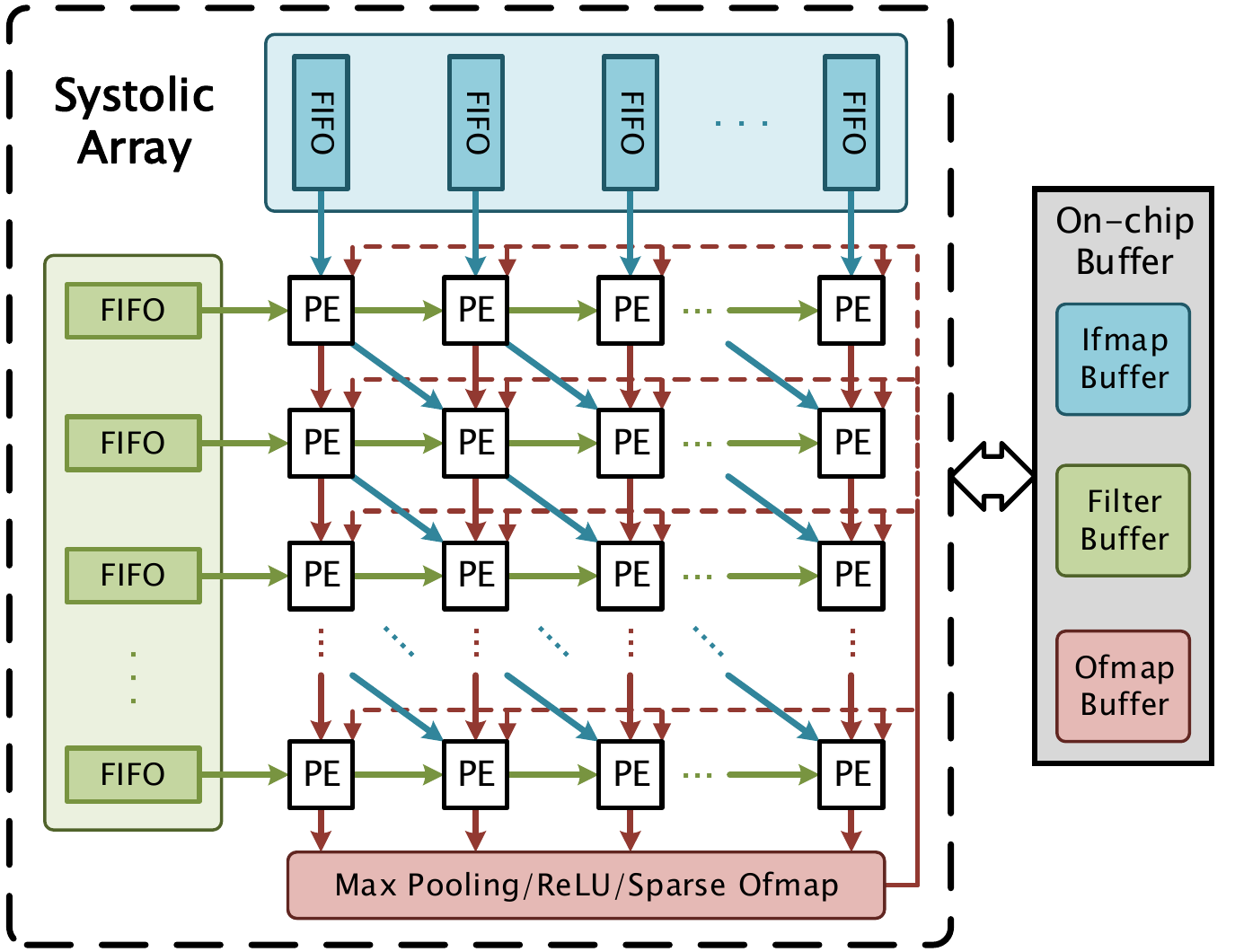}
\caption{Systolic array hardware structure for neural networks~\cite{17}.}
\label{fig:multideploy}
\end{figure}

 The neural networks are highly self-adaptive and reliable.
 By applying DNN and CNN in the baseband, we can achieve performance enhancements as long as
 a uniform hardware implementation framework. Actually, the core operation for CNN is the convolution,
 while the core of DNN is the multiplication of the two-dimensional matrices. We notice that
 the systolic architecture can realize both of the operations.
 Figure \ref{fig:multideploy} illustrates a reconfigurable systolic architecture designed for
 accelerating convolutional neural network~\cite{17}. It can
 be seen that the systolic architecture is regular and scalable, which supports different CNNs and DNNs.
 This motivates us to explore the possibilities of reusing the same hardware architecture to realize both
 5G and DL algorithms.

\begin{figure}[!t]
\centering
\includegraphics[width=6in]{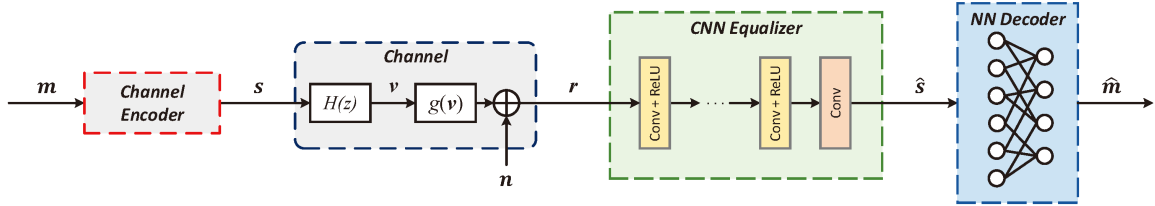}
\caption{Architecture of a receiver including neural network equalizer and decoder~\cite{20}.}
\label{fig:joint}
\end{figure}

The authors of \cite{20} point out that in a system formed with the channel encoder, the channel,
the channel equalizer and the decoder (as shown in Figure \ref{fig:joint}), the equalizer and the decoder
in the receiver can be implemented with a CNN and a DNN, respectively.
The associated AI accelerator can be jointly realized by two strategies:

(1) The uniform architecture: The overall receiver can be folded into one uniform processor to save the hardware area.
This processor firstly works as a CNN-based equalizer with the input signals from the channel.
The output of the CNN will be saved at this point.
The processor will then function as a DNN-based decoder, for which the saved output from the CNN will
serve as the input. The decoding results will be the final output.
(2) The cascade architecture: Two processors will be cascaded directly to construct the receiver,
one being the CNN-based equalizer while the other being the DNN-based decoder.
This architecture has more hardware consumption, but achieves higher throughput rate.

Overall, the AI-based uniform accelerator is more flexible
for the hardware implementation, hence can achieve various system requirements.

\subsection{AI for physical layer: DNN-based end-to-end communication}

As mentioned above, AI, especially DNN, has been applied to different function blocks in the physical layer,
e.g., modulation recognition~\cite{42}, polar codes decoder~\cite{19} and MIMO detection~\cite{41}.
For the joint optimization of two or more blocks, AI algorithms also achieve success,
e.g., the aforementioned joint optimization of channel equalizer and channel decoder proposed in~\cite{20}.
However, optimizing each of the blocks individually cannot guarantee the optimization for the whole physical layer
communication~\cite{43}. In the viewpoint of the whole end-to-end communication system,
the intuitive connection of different AI modules may bring extra computational cost for both training and online tasks.
Hence, a joint optimization method for the end-to-end system is in need.

\begin{figure}
\centering
\includegraphics[width=4in]{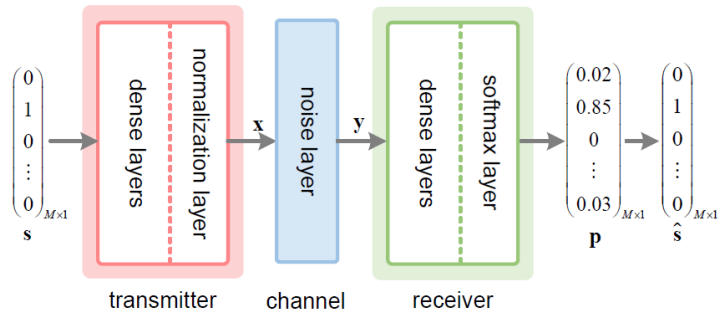}
\caption{A simple autoencoder for an end-to-end communication system~\cite{43}.}
\label{fig:autoencoder}
\end{figure}

\cite{44} proposes an autoencoder-based end-to-end system optimization method,
in which the communication system is recast as an end-to-end reconstruction optimization problem,
and a novel concept, the autoencoder, is introduced to serve as a simplified representation of the system.
The autoencoder is a type of ANN. It aims at learning a representation (encoding) for a set of data
in an unsupervised manner, which will be capable to reconstruct compressed inputs at the output layer.
In the proposed approach in \cite{44}, the end-to-end system is simply represented with three blocks,
the transmitter, the receiver and the channel.
The transmitter and the receiver are both represented as fully connected DNNs.
A normalization layer is connected to the transmitter to guarantee the energy constraints,
whereas a softmax activation layer is placed before the receiver to output soft decisions for the received information.
The AWGN channel in between is represented as a simple noise layer with a certain variance.
The resulted autoencoder has a structure as shown in Figure \ref{fig:autoencoder}.
This autoencoder is trained based on bit error rate (BER) or block error rate (BLER).
After training, it will be able to reconstruct the transmitted signals based on the received signals.

\begin{figure}
\centering
\includegraphics[width=4.5in]{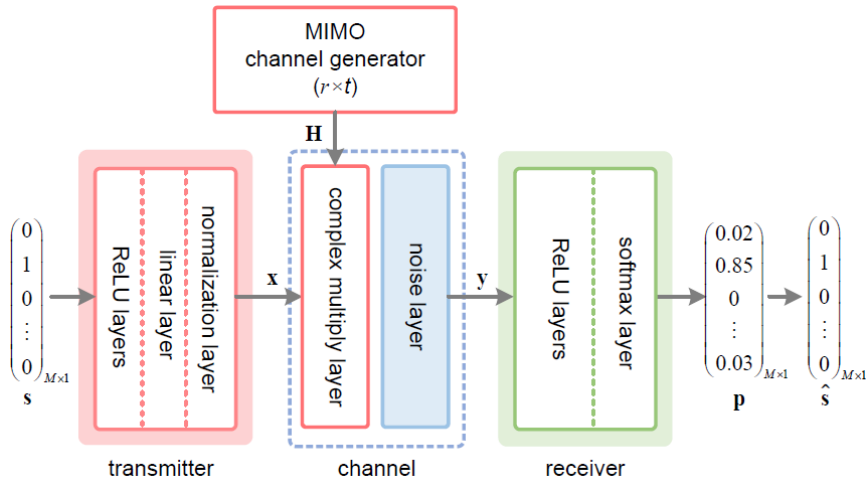}
\caption{A general MIMO channel autoencoder architecture~\cite{43}.}
\label{fig:aemimo}
\end{figure}

The autoencoder is a novel concept different to all the traditional, conventional methods.
DNNs are utilized to represent the whole end-to-end system without considering the specific models
in each traditional function block.
Hence, in scenarios that is too complicated to model,
autoencoder can be a good solution to ``learn" these scenarios and optimize the performance.
This scheme is extended to a multi-user scenario with interfering channels in \cite{44},
and is further extended to MIMO in \cite{45} by adding a module for the channel matrix as shown in Figure \ref{fig:aemimo}.
Simulation results presented in \cite{44,45} show that the autoencoder can ``learn" different scenarios
with various CSI and number of antennas and achieves enhanced BER performance.

\section{Conclusions}
\label{sec:conclustion}

5G makes a significant breakthrough in the traditional mobile communication system. While enhancing the service
capability of the traditional mobile networks, it further evolves to support the applications of IoT in various
fields including business, manufacturing, health care and transportation. Hence, 5G is becoming the basic technology
for the future IoT that connects and operates the whole society. Aiming at supporting differentiated applications
with a uniform technical framework, 5G is facing enormous challenges.
With the revive and the rapid development in recent years, AI is rising to these challenges.
It is a potential solution to the problems in the 5G era, which will lead to
a revolution in the concepts and capabilities of the communication systems.

Many researches have already been done for applying AI in 5G. In this paper, instead of trying to review all the
existing literatures, we focus on clarifying the promising research directions with the greatest potential.
By further efforts in these research directions, 5G can be anticipated to achieve significantly better performance and more convenient implementations compared to the traditional communication systems.
With the inspiring research paradigms introduced in this paper, we are looking forward to the remarkable achievements of AI in 5G in the near future.

\Acknowledgements{This work was supported by National Natural Science Foundation of China (Grant Nos. 61501116 and 61221002).}





\end{document}